\documentclass[amsmath,amssymb,aps,nopacs,prb,floatfix,preprint]{revtex4}
\usepackage{stmaryrd}
\usepackage{dcolumn}
\usepackage{bm}
\usepackage{color}
\usepackage{multirow}
\usepackage{times}
\usepackage{graphicx}

\renewcommand{\Im}{\mathrm{Im}}

\newcommand{\Tr}{\mathrm{Tr}}

\newcommand{\Tmat}{\ensuremath{\mathcal{T}}}

\newcommand{\vc}[1]{\ensuremath{\boldsymbol{#1}}}

\newcommand{\vk}{\ensuremath{\vc{k}}}

\newcommand{\BT}{B\MakeLowercase{i}$_2$T\MakeLowercase{e}$_3$}
\newcommand{\BS}{B\MakeLowercase{i}$_2$S\MakeLowercase{e}$_3$}
\newcommand{\Tevac}{{$\mathrm{Vac}_{\mathrm{Te}}$}}

\newcommand{\cita}{\cite}
\newcommand{\lbl}{\label}

\begin{document}

\title{Lifetime and surface-to-bulk scattering off vacancies of the topological surface state in the three-dimensional strong topological insulators \BT\ and \BS}

\author{Philipp R\"u{\ss}mann} 
\email{P.Ruessmann@fz-juelich.de}
\affiliation{Peter Gr\"unberg Institut and Institute for Advanced Simulation, 
	Forschungszentrum J\"ulich and JARA, D-52425 J\"ulich, Germany}
\author{Phivos Mavropoulos}\email{Ph.Mavropoulos@fz-juelich.de} 
\author{Stefan Bl\"ugel} 
\affiliation{Peter Gr\"unberg Institut and Institute for Advanced Simulation, 
	Forschungszentrum J\"ulich and JARA, D-52425 J\"ulich, Germany}

\begin{abstract}
We analyze the finite lifetimes of the topologically protected electrons in the surface state of \BT\ and \BS\ due to elastic scattering off surface vacancies and as a function of  energy. The scattering rates are decomposed into surface-to-surface and surface-to-bulk contributions, giving us new fundamental insights into the scattering properties of the topological surface states (TSS). 
If the number of possible final bulk states is much larger than the number of final surface states, then the surface-to-bulk contribution is of importance, otherwise the surface-to-surface contribution dominates. Additionally, we find defect resonances that have a significant impact on the scattering properties of the TSS. They can strongly change the lifetime of the surface state to vary between tens of fs to ps at surface defect concentrations of 1~at\%.	
\end{abstract}

\maketitle

\section{Introduction}

Typical samples of the binary strong topological insulators of the \BT\ and \BS\ family contain native defects so that the Fermi level is at a position where bulk states coexist with the topological surface states \cita{Chen2009}. 
This raises the question on the lifetime of surface state electrons in the presence of bulk states or, more generally, on the energy-dependence of the scattering rate at varying energies in the conduction band,  the bulk band gap, and the valence band. The importance of these questions is obvious as the scattering features of the TSS play a major role in surface transport properties and in the wished long coherence length for topological-insulator based applications.

Nowadays, different strategies exist to pin the position of the Fermi level into the band gap in order to suppress the number of bulk states at the Fermi level. The task can be achieved by compositional tuning (see for example Refs.~\cita{Taskin2011, Kellner2015}) or by compensating individual doping trends in forming topological $p$-$n$-junctions \cita{Eschbach2015}. Nevertheless, typical transport  measurements in this class of materials still fight against contributions from bulk states, as opposed to the desired surface-dominated transport,  and a cumbersome disentanglement of surface and bulk contributions is needed (e.g. Refs.~\cita{Taskin2011, Hinsche2015, deCastro2016}). 

In addition to transport measurements, two-photon photo-emission spectroscopy
allows to get information about the dynamics of excited electrons in topological insulators \cita{Sobota2012, Niesner2014, Cacho2015, Kuroda2016, Kuroda2017}. In \BS, for example, a persisting population for longer than 10\,ps \cita{Sobota2012} has been found, which was attributed to a dominant contribution to the relaxation dynamics from elastic scattering of excited electrons between surface and bulk states. Even longer lifetimes of the surface state exceeding 4\,ps were reported recently  \cita{Cacho2015}. However, these lifetimes are the result of a complicated interplay between elastic and inelastic scattering and only recently in {Sb$_2$Te$_3$} elastic
lifetimes of 300\,fs-2.5\,ps were found using mid-IR pump pulses, that allow to excite direct transitions from occupied to unoccupied parts within the topological surface state \cita{Kuroda2016, Kuroda2017}. 

So far, from a theoretical point of view, the question of surface-to-bulk defect scattering has only been treated based on simple models \cita{Saha2014}. Here, an ab-initio treatment will be presented, that gives valuable insights into the energy-dependence of the scattering properties of the TSS in strong topological insulators and allows to draw conclusions about their robustness against surface impurities. As defects we considered a single Te (Se) vacancy in the surface layer  of \BT\ (\BS). This particular defect has been chosen because of its very frequent occurrence in this class of materials \cita{Sessi2014} and because vacancies are suspected to be the reason for n-type doping in \BT\ \cita{Oostinga2013, Bathon2016}. 
The main difference in the band structure between \BT\ and \BS\ is that the warping term, shaping the constant-energy contour of the TSS  from a circular form to a star-like form, is more pronounced  in \BT. A comparison between the two systems will therefore give information on the importance of the warping term in the TSS scattering properties.

The work presented here aims at answering the following questions: what is the energy-dependence of the scattering rates off non-magnetic defects and how is it related to the defect properties, e.g. the defect density of states? Furthermore, if the TSS coexists with bulk states, what are surface-to-bulk scattering rates and how do they relate to the surface-to-surface scattering rates? We find that three factors are of major importance in answering these questions for non-magnetic defects: (i) the defect density of states, (ii) the number of available final states in the scattering process and (iii) the localization of the wavefunctions, in particular their spatial overlap with the impurity potential. The comparison of \BT\ with \BS\ allows us to draw the conclusion that the warping of the constant energy contour, which is stronger in \BT\ than in \BS, does not significantly alter the total scattering rate off impurities. This result suggests that our findings are of general importance for strong three-dimensional topological insulators of this family of compounds, irrespective of the exact shape of the Fermi surface.

\section{Method of calculation}

\subsection{Lifetime of the TSS and decomposition of scattering rates}
Our investigations are based on the calculation of the  elastic scattering rate caused by the defect potential, 
\begin{equation}
P_{\vk'\vk}= \frac{2\pi}{\hbar} Nc |\Tmat_{\vk'\vk}|^2\delta(E_{\vk}-E_{\vk'})
\end{equation}
between initial ($\vk$) and final ($\vk'$) states. Here, $c$ is the defect concentration and $N$ the number of atoms in the surface layer, i.e., $Nc$ is the total number of defects. $\Tmat_{\vk'\vk}$ is the \Tmat-matrix element of the defect. We define the $\vk$-dependent relaxation rate as an inverse lifetime, $\tau_{\vk}^{-1}=\sum_{\vk'}P_{\vk'\vk}$, and the relaxation rate averaged over the topological surface states at $E$, $\tau^{-1}_{\mathrm{s}}(E)=n_{\mathrm{s}}^{-1}(E)\sum_{\vk}^{\mathrm{s}(E)}\tau_{\vk}^{-1}$, normalized to the density of surface states $n_{\mathrm{s}}(E)$. We also adopt the decomposition of $\tau^{-1}_{\mathrm{s}}(E)$ in contributions from scattering into bulk end-states and into TSS end-states:
\begin{eqnarray} \tau^{-1}_{\mathrm{s}}(E)&=&\tau^{-1}_{\mathrm{sb}}(E)+\tau^{-1}_{\mathrm{ss}}(E)
 \lbl{eq:decomposition}\\
&=&n^{-1}_{\mathrm{s}}(E)\sum_{\vk}^{\mathrm{s}(E)} \left[
\sum_{\vk'}^{\mathrm{b}(E)}P_{\vk'\vk} + \sum_{\vk'}^{\mathrm{s}(E)}P_{\vk'\vk}
\right] \nonumber
\end{eqnarray}
(with obvious correspondence between the terms on the r.h.s. of the equation defining the two contributions).
Here,  $\mathrm{s}(E)$ and $\mathrm{b}(E)$ stand for the set of surface states and bulk states, respectively, at energy $E$. Finally, we denote the bulk density of states per quintuple layer as $n_{\mathrm{b}}(E)$.

Our description addresses the low-temperature regime and energies close to the Fermi level, where the impurity scattering dominates over the electron-electron scattering rate that behaves as $\tau_{e-e}^{-1}\propto(E-E_F)^2$ according to Landau's Fermi-liquid theory and where the phonons  are frozen out.  In fact, ab-initio calculations of the electron phonon coupling constant for \BT\ and \BS\ \cita{Heid2017} showed that the coupling strength is quite weak, which highlights the importance of our calculations for transport properties.

\subsection{Electronic structure of the host and defect}

The \BT\ and \BS\ surfaces were simulated by considering
films of six quintuple-layers thickness. The atomic positions followed the experimental bulk lattice structure (see Ref.\ \onlinecite{Nakajima1963} for \BT\ and Ref.\ \onlinecite{Wyckhoff1964} for \BS). The electronic structure was calculated  within the local density approximation (LDA) \cita{Vosko1980} to density functional theory by employing the full-potential relativistic Korringa-Kohn-Rostoker Green function method (KKR) \cita{Bauer2013,Ebert2011} with exact description of the atomic cells \cita{Stefanou1990,Stefanou1991} and an $\ell_{\rm max}=3$ cutoff in the angular momentum expansion. The position of the Te-vacancy for the \BT\ surface (Se-vacancy for the \BS-surface) is shown in Figure~\ref{fig:ch5setup}(c). The defects, together with a charge-screening cluster comprising the first shell of neighboring atoms, was embedded self-consistently using the Dyson equation in the KKR method \cita{Bauer2013}. The perturbed potential and density were also calculated within the LDA. The \Tmat-matrix and scattering rates were calculated in the KKR representation as described elsewhere \cita{Long2014, Heers2012}. Structural relaxations around the vacancy can be important in the calculation of the vacancy-formation energy. However, concerning the trends of the scattering rate, they are expected to play only a minor role, and were therefore neglected.

\section{Energy dependent scattering rates of the TSS in \BT}

\begin{figure}[htbp]
	\centering
	\includegraphics[width=0.80\linewidth]{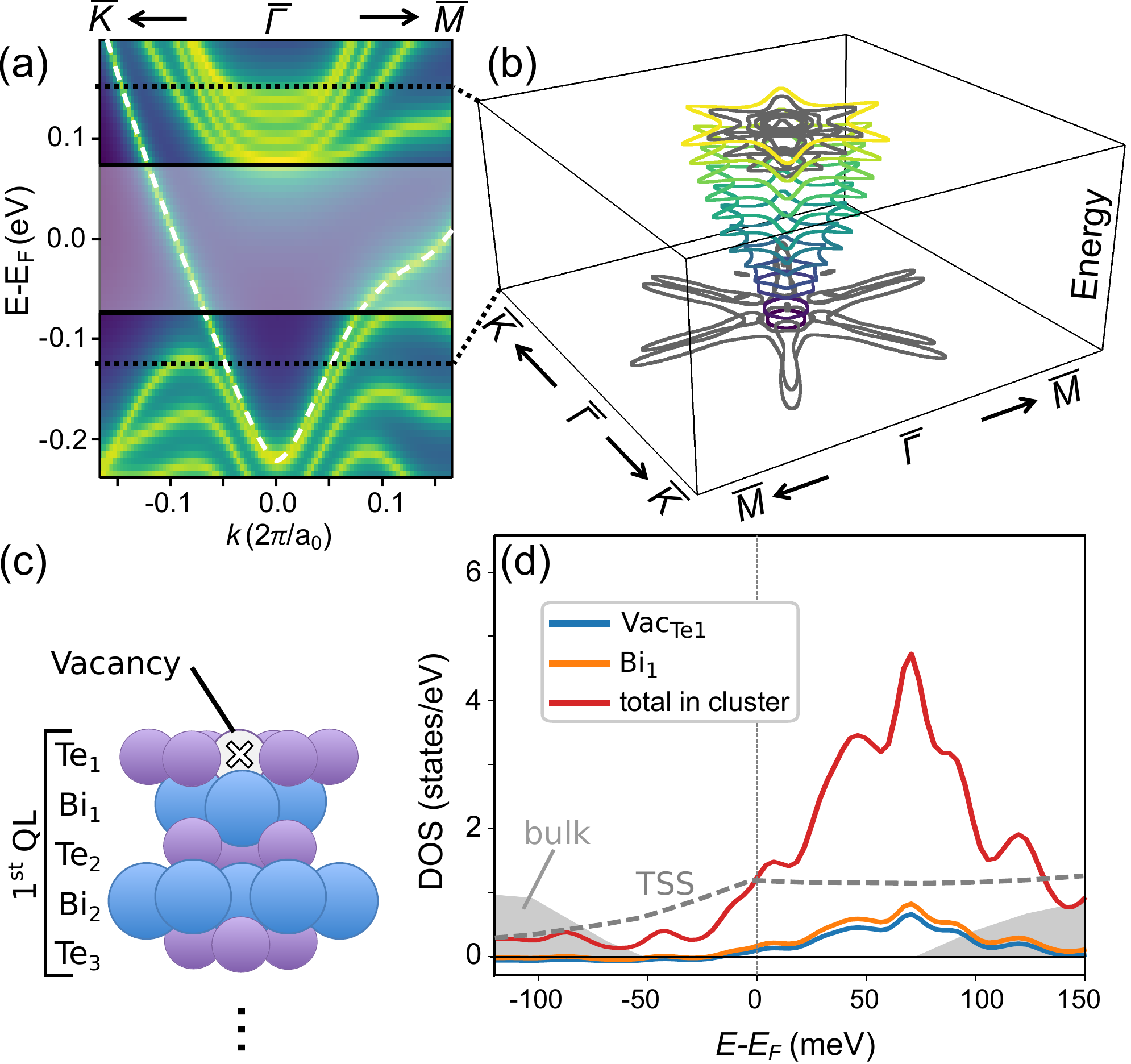}
	\caption{Setup of the calculation. 
		(a) Band structure of a 6QL \BT\ film, where the shaded area between the two horizontal black lines highlights the region of the bulk band gap and the black dotted lines indicate the energy range for which the scattering rates are analyzed. 
		(b) 3D plot of the dispersion where the bulk bands are shown in gray and the TSS (white dashed line in (a)) in a color code representing the energy.
		(c) Side view of the vacancy position in the surface quintuple layer.
		(d) Density of states (DOS) of bulk-states corresponding to one quintuple layer (shaded area), of the TSS (dashed line), and of the vacancy defect in the first Te-layer of \BT, including the total DOS (sum over full impurity cluster containing the vacancy and the surrounding atoms, red), as well as the contributions from the vacancy position ($\mathrm{Vac}_{\mathrm{Te}}$, blue) and the neighboring Bi-atoms (Bi$_1$, orange). The resonance seen in $\mathrm{Bi}_1$ is associated to Bi dangling bonds projecting into the vacancy site and is absent without the defect.}
	\lbl{fig:ch5setup}
\end{figure}

Starting with the \BT\ system, we show the band structure in Figure~\ref{fig:ch5setup}(a,b) and the density of states of the bulk (one quintuple layer), the TSS, and impurity, in Figure~\ref{fig:ch5setup}(d). 
The vacancy shows a resonance, with a maximum of the DOS at $E-E_F\approx75\mathrm{\,meV}$, which is absent in the ideal surface. The resonance results from dangling bonds of the adjacent Bi$_1$ position as they project into the  vacancy (see the corresponding DOS in Figure~\ref{fig:ch5setup}d).

The scattering rate $P_{\vk'\vk}$
shows expected trends in terms of weighted forward scattering and reduced backscattering, with vanishing exact time-reverse scattering, as depicted in Figure \ref{fig:ch5scattratesBT}(a) for an initial state on the TSS marked by a circle. This statement  even holds at energies outside the bulk band gap, where the TSS coexists with bulk states from the valence or conduction band. 


\begin{figure}[htbp]
\centering
\includegraphics[width=0.9\linewidth]{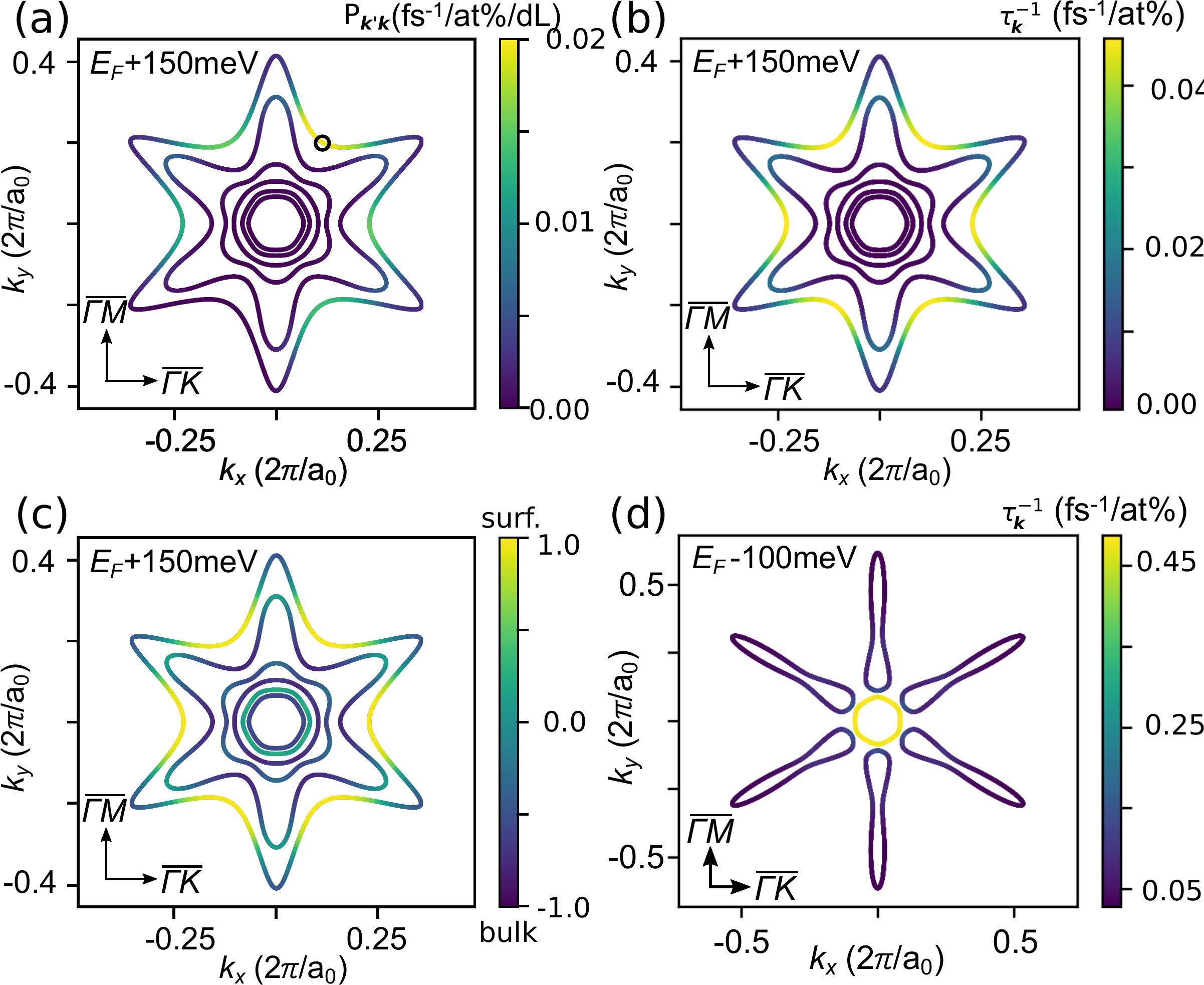}
\caption{Scattering rates of topological surface state off a surface vacancy in \BT.  (a) $P_{\vc{k}'\vc{k}}$ for an initial state $\vk$ marked by a small circle on the TSS (outermost loop) at an energy of $E_F+150\mathrm{\,meV}$, where also conduction-band bulk states are present, and end-states $\vk'$ running over the whole constant-energy surface. The color code indicates the rate per line segment of the constant-energy contour. Exact time-reverse scattering is absent, while near-time-reverse scattering is very weak. Forward scattering is promoted. (b) 
Inverse relaxation time, $\tau^{-1}_{\vk}$, of the states on the constant energy contour at $E_F+150\mathrm{\,meV}$. 
The topological surface states (outermost, star-like line) is scattered more effectively (yellow regions) than the bulk-derived states (inner lines) of the conduction band. A strong anisotropy in the scattering rates is seen that leads to smaller relaxation times of states in the valleys in 
$\overline{\Gamma K}$ as opposed to the states at the tips of the star in $\overline{\Gamma M}$. (c) Localization function of the states at
 $E_F+150\mathrm{\,meV}$
 in the surface quintuple layer vs.\ the 3rd quintuple layer. (d) Same as in (b), but at the top of the valence band ($E_F-100\mathrm{\,meV}$) where the TSS is the innermost loop.}
	\lbl{fig:ch5scattratesBT}
\end{figure}

The $\vk$-dependent relaxation rate, $\tau^{-1}_{\vk}(E)$, is shown in Figure~\ref{fig:ch5scattratesBT}(b) at the conduction band edge ($E=E_F+150\mathrm{\,meV}$), where the TSS (outermost contour) coexists with bulk states (inner contours), and in Figure~\ref{fig:ch5scattratesBT}(d) at the valence band edge ($E-E_F=-100\mathrm{\,meV}$), where the TSS is the innermost ring and the bulk states are the elongated closed loops in the $\overline{\Gamma M}$ directions. In both cases, mostly the TSS is scattered (as the color code indicates), while the bulk states remain comparatively unaffected by the defect. The reason for this difference is the large overlap of the surface-localized TSS with the surface impurity, compared with the low overlap of the delocalized bulk states. In the limit of large film thickness, and since there are no surface resonances in this energy range, the overlap of the bulk states with the impurity will tend to vanish together with the scattering rate. 

Importantly,  we witness a correlation between the warping of the TSS and the $\vk$-resolved scattering rate. As we see in Figure~\ref{fig:ch5scattratesBT}(a), $\tau^{-1}_{\vk}(E)$ is highly anisotropic on the TSS, with higher values at the valleys ($\overline{\Gamma K}$ directions) and strongly reduced values by almost an order of magnitude (factor 6.8) at the tips of the constant energy contour ($\overline{\Gamma M}$ direction). The anisotropy was also seen in angular-resolved photoemission experiments \cita{SanchezBarriga2014} where the line-width of the spectral density can be related to the scattering rate of the states. The effect is consistent with the fact that the states at the tips of the warped contour are more delocalized, projecting into the bulk, as compared with the states in the valleys that have a larger amplitude in the surface layer. The localization is demonstrated in Figure \ref{fig:ch5scattratesBT}(c), where we plot  a localization function 
\begin{equation}
g_{\vk}=\frac{n_{\vk}(\text{1st QL})-n_{\vk}(\text{3rd QL})}{n_{\vk}(\text{1st QL})+n_{\vk}(\text{3rd QL})}
\lbl{eq:loc}
\end{equation}
where $n_{\vk}(\text{1stQL})$ the spatial density of the state at $\vk$ integrated in the 1st (surface) quintuple layer and $n_{\vk}(\text{3rd QL})$ is the same for the 3rd quintuple layer (middle of the film). The function takes values $-1\leq g_{\vk}\leq 1$, with positive values corresponding to surface localization and negative values to bulk localization. Evidently, the valleys of the TSS are characterized by an increased localization (yellow color), compared to the tips of the TSS (green).

The anisotropy of $\tau_{\vk}$ is absent close to the conduction band edge [Figure~\ref{fig:ch5scattratesBT}(d)] where there is no warping.
These results suggest that the widely used constant relaxation-time approximation in the calculation of transport properties may be insufficient in cases of surface-defect scattering, if the Fermi contour is warped.


The energy dependence of the scattering rate is shown in Figure~\ref{fig:ch5scattratesBT}(b). 
The dominant feature is the very strong variation of  $\tau_{\mathrm{s}}^{-1}(E)$, spreading over  almost two orders of magnitude, between  $\tau_{\mathrm{s}}\approx66\mathrm{\,fs\,at\%} $ for $E\approx E_F+75\mathrm{\,meV}$  and  $\tau_{\mathrm{s}}\approx3.3\mathrm{\,ps\,at\%} $ around $E\approx E_F-100\mathrm{\,meV}$. This peaking form is characteristic of resonant scattering. It follows the resonance seen in the density of states of the impurity (cf. Figure~\ref{fig:ch5setup}(d)), highlighting the 
well-known generalized Friedel sum rule, $n_{\rm imp}(E)-n_{\rm host}(E) = \frac{1}{\pi}\frac{d}{dE} \Im \Tr \ln \Tmat(E)$.

\begin{figure}[htbp]
	\centering
    \includegraphics[width=0.99\linewidth]{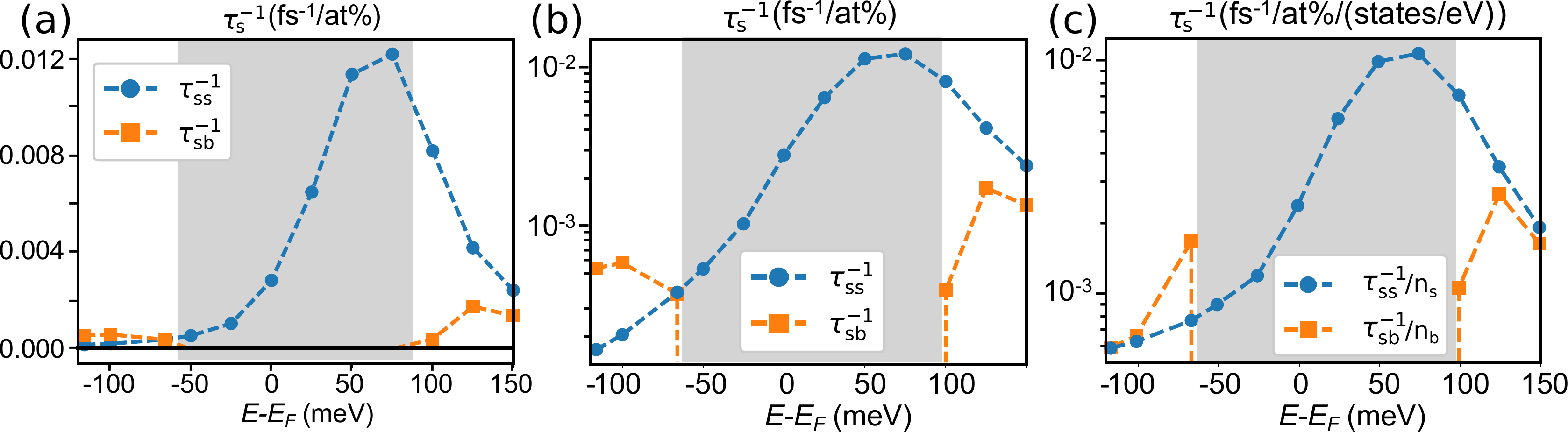}
	\caption{(a) Energy-dependence of the surface-to-surface ($\tau_{\mathrm{ss}}^{-1}$) and surface-to-bulk ($\tau_{\mathrm{sb}}^{-1}$) contributions to the scattering rate off a surface Te-vacancy on \BT. A different behavior is observed for $E$ in the valence band, where $\tau_{\mathrm{ss}}^{-1}<\tau_{\mathrm{sb}}^{-1}$, as compared to the conduction band, where $\tau_{\mathrm{ss}}^{-1}>\tau_{\mathrm{sb}}^{-1}$. The resonant scattering off the \Tevac\ impurity (cf. DOS in Figure~\ref{fig:ch5setup}(d)) leads to strong surface-to-surface scattering rates with a maximum around $E-E_F\approx75\mathrm{\,meV}$. (b) same as in (a), but on a logarithmic scale. (c) Rescaled scattering rates showing the importance of the number of possible final states for the scattering of the TSS to bulk states. The gray shaded areas correspond to the position of the bulk band gap.}
	\lbl{fig:ch5phasespace}
\end{figure}

The decomposition of the scattering rate, presented in Figure \ref{fig:ch5phasespace}(a) and in a logarithmic scale  in Figure~\ref{fig:ch5phasespace}(b),
 shows a qualitatively different behavior   in the conduction band ($\tau_{\mathrm{ss}}^{-1}>\tau_{\mathrm{sb}}^{-1}$) compared to the valence band ($\tau_{\mathrm{ss}}^{-1}<\tau_{\mathrm{sb}}^{-1}$). 
The latter observation seemingly contradicts the interpretation in terms of the overlap between end-states and defect potential:  surface end-states, that are more localized at the surface, should in principle produce larger scattering rates than bulk end-states. However, we must also consider the phase space of available  end-states at the valence band. Firstly, the constant-energy contour of the TSS is closing to the Dirac point, and secondly, the bulk DOS is enhanced because the  constant-energy contour at the onset of the valence band contains elongated bulk hole pockets (Figures \ref{fig:ch5setup}(b) and \ref{fig:ch5scattratesBT}(b)). Consequently, the phase-space ratio of bulk final states to TSS final states, $n_{\mathrm{b}}(E)/n_{\mathrm{s}}(E)$,  is larger in the valence band as compared to the onset of the conduction band and plays a central role in the relative contribution to the scattering rate. [$n_{\mathrm{b}}(E)$ and $n_{\mathrm{s}}(E)$ are shown in Figure \ref{fig:ch5setup}(d).]

The phase space of the final states can be factored out by re-normalizing the surface-to-surface and surface-to-bulk rates with respect to the density of final states, plotting $\tau_{\mathrm{ss}}^{-1}(E) \,n^{-1}_\mathrm{s}(E)$ and $\tau_{\mathrm{sb}}^{-1}(E)\, n^{-1}_\mathrm{b}(E)$. This is shown in Figure~\ref{fig:ch5phasespace}(c), where we see that the differences between surface-to-surface and surface-to-bulk contributions are not as pronounced in the re-normalized rates as they are in the ``normal'' rates [Figure~\ref{fig:ch5phasespace}(b)]. In particular, in the valence band, $\tau_{\mathrm{ss}}^{-1}(E) \,n^{-1}_\mathrm{s}(E) < \tau_{\mathrm{sb}}^{-1}(E)\, n^{-1}_\mathrm{b}(E)$, while in the conduction band $\tau_{\mathrm{ss}}^{-1}(E) \,n^{-1}_\mathrm{s}(E) > \tau_{\mathrm{sb}}^{-1}(E)\, n^{-1}_\mathrm{b}(E)$.

With respect to the finite film thickness considered in the calculations, we should comment that the quantities $n_\mathrm{b}(E)$ and $\tau_{\mathrm{sb}}^{-1}(E)$ both saturate in the limit of large thicknesses. This is because, on the one hand, the overlap of the bulk states with the defect becomes smaller, while, on the other hand, the number of bulk states increases. These two contributions counter-act each other and lead to a saturation, as was explained in more detail in Ref. \onlinecite{Heers2012}. Thus our result at a thickness of 6 quintuple layers is close to the half-infinite crystal limit.
Our findings underline the importance of the final states' phase space, which amplifies the surface-to-bulk scattering rate, in agreement with the model-based calculations by Saha et al. \cita{Saha2014}. 

Our observation certainly holds true for the energy range that was studied here. However, it could be different if surface-resonances occur in the band structure, where bulk states have a significant weight at the surface. In \BS, spin-polarized surface resonances were recently found in the unoccupied band structure approximately 0.5eV above the conduction band edge \cita{Cacho2015}. The surface-localization of these states and their spin-polarization could then have a significant impact on possible scattering events of the TSS.

\section{The influence of the warping strength: comparison to \BS}

Finally, the influence of the warping strength in the surface states scattering will be analyzed further by considering the case of scattering off a Se-vacancy in \BS, since the TSS warping is much smaller in this material. The results are summarized in Figure~\ref{fig:ch5BSresults}. The reduced warping in \BS\ can be seen in the band structure shown in Figure~\ref{fig:ch5BSresults}(a) and (b). The density of states of the Se-vacancy defect embedded into the surface shows a similar behavior as the Te-vacancy in \BT\ with a resonance in the DOS lying within the bulk band gap (cf. Figure~\ref{fig:ch5BSresults}(c)), around $E-E_\mathrm{D}\approx150\mathrm{\,meV}$ ($E_\mathrm{D}$ is the energy of the Dirac point). The energy-dependent $\tau_{\mathrm{ss}}^{-1}$ and $\tau_{\mathrm{sb}}^{-1}$ for the Se-vacancy are shown in Figure~\ref{fig:ch5BSresults}(c) in a semi-logarithmic scale. A very sharp resonance in the surface-to-surface scattering rate is found which exactly coincides with the impurity DOS peak of the vacancy defect, as expected. We conclude that the warping, which becomes important only at higher energies, very close to the conduction band (cf. Figure~\ref{fig:ch5BSresults}(b)), has no significance for the appearance of the resonance.

\begin{figure}[htbp]
	\centering
		\includegraphics[width=0.9\linewidth]{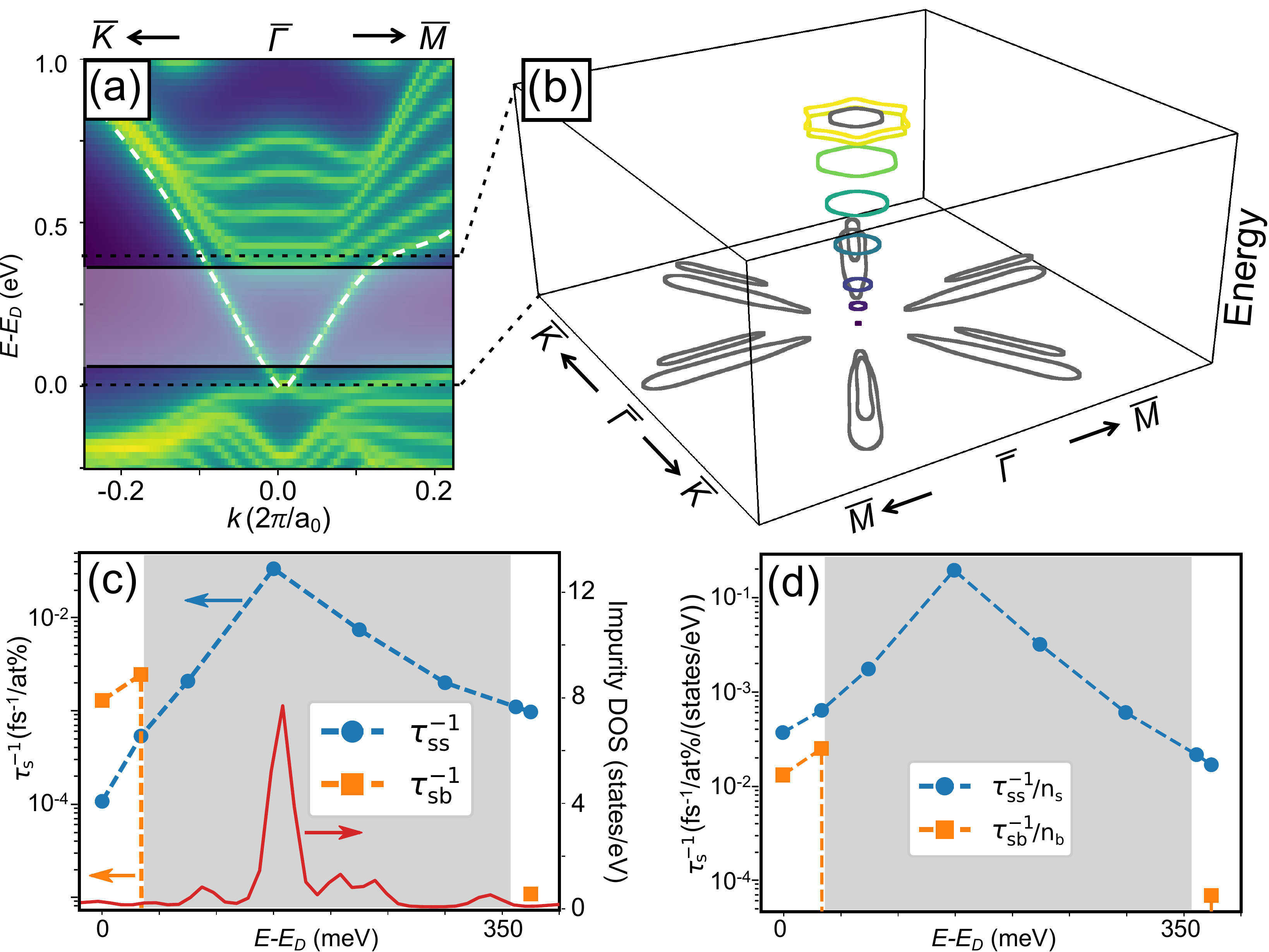}
	\caption{Scattering rates off a surface Se-vacancy in \BS. (a) Band structure where the dispersion of the topological surface state is highlighted with white dashed lines. The light shaded area (between the black horizontal lines) denotes the bulk band gap and the black dotted lines indicate the range in which the scattering rate and lifetimes have been analyzed. $E_\mathrm{D}$ is the energy of the Dirac point.
	(b) 3D plot of the dispersion where the bulk bands are shown in gray and the TSS (white dashed line in (a)) in a color code representing the energy.
	(c) Total density of states of the Se-vacancy (red, full line) and energy-dependence of the surface-to-surface ($\tau_{\mathrm{ss}}^{-1}$) and surface-to-bulk ($\tau_{\mathrm{sb}}^{-1}$) scattering rates for a Se-vacancy defect. (d) The scattering rates re-normalized to the density of final states, showing the importance of the available scattering phase space. The gray shaded areas in (c,d) correspond to the position of the bulk band gap and the energies are given with respect $E_\mathrm{D}$.}
	\lbl{fig:ch5BSresults}
\end{figure}

Similarly to the case of \BT, within the valence band the surface-to-bulk scattering rate is larger than the surface-to-surface rate, while the reversed situation occurs at the onset of the conduction band. To highlight the influence of the final state phase space we again re-normalize with respect to the density of final states, depicted in Figure~\ref{fig:ch5BSresults}(d). The re-normalized rates also show an analogous behavior as in \BT.
The strong variation of the scattering rates due to resonant scattering off the impurity leads to lifetimes that vary between $\tau_{\mathrm{s}}\approx165\mathrm{\,fs\,at\%}$ at $E-E_\mathrm{D}\approx150\mathrm{\,meV}$ and $\tau_{\mathrm{s}}\approx6.5\mathrm{\,ps\,at\%}$ at $E=E_\mathrm{D}$. 

We can thus summarize that the warping is not significant for the average scattering rates, since \BS\ shows an analogous behavior as  \BT. The defect resonance plays a dominant role in the energy dependence of the rates. The TSS mainly scatters to other states within the TSS at the onset of the conduction band, while the surface-to-bulk scattering prevails in the valence band due to the much larger density of bulk states.

\section{Summary and conclusions}

The energy-dependence of scattering rates (inverse lifetimes) of the topological surface state off surface vacancies on \BT\ and \BS\ has been studied from first principles. These defects occur frequently and are important for intrinsic doping as well as transport properties. 

We located dominant features of resonant scattering of the TSS, reflecting the defects' density of states and caused by dangling bonds. The importance of resonant scattering in metals is well known, including trivial surface states as e.g.\ the (111) surface states of the noble metals \cita{Heers2012, Ruessmann2014}. Here we have seen that the TSS are just as susceptible to resonant scattering, with the difference that the scattering is mostly forward-directed and exact or near backscattering is absent or strongly suppressed, respectively. 
For a qualitatively and quantitatively realistic description of the scattering, ab-initio calculations are therefore of paramount importance, since they have the power to predict the resonance position by the self-consistent charge-density screening of the defect. 

At energies where the TSS coexists with bulk states, we find a different behavior in the valence versus in the conduction band. In the valence band, scattering into bulk states is dominant, promoted mainly by of the overwhelming density of available bulk end-states in comparison to the TSS. Things are opposite at the onset of the conduction band, where we find larger surface-to-surface scattering rates. Here, an electron that has potentially been prepared in the TSS (e.g. in a surface state dominated transport experiment or in STM) stays mostly within the TSS after scattering. This is consistent with recent quantum transport measurements in \BS, where the phase coherence length and a strong focusing in forward scattering direction was even seen for strongly disordered samples \cita{Dufouleur2013, Dufouleur2016}. We expect that this behavior will hold also in other TIs of the same family and other impurities, as long as the surface band structure does not host surface resonances.


The warped contour of the surface state in \BT\ was found not to influence the trends of the average relaxation rate, but to have a large impact on the anisotropy of the relaxation rate with respect to $\vk$, in agreement with experiments \cita{SanchezBarriga2014}. This is a consequence of the different localization of the TSS at different states on the constant energy contour. The constant relaxation time approximation might therefore not be sufficient for the description of transport phenomena in case of surface defects.

\section*{Acknowledgments}

We gratefully acknowledge financial support from the DFG (SPP-1666, Project No. MA 4637/3-1) and from the VITI project of the Helmholtz Association as well as computational support from the JARA-HPC Supercomputing Centre at the RWTH Aachen University.


\end{document}